\documentclass[a4paper,11pt]{article}

\usepackage{graphicx}
\usepackage{natbib}
\usepackage{here}
\usepackage{setspace}
\usepackage{multirow}
\usepackage{amsmath, amssymb}
\usepackage{bm}
\usepackage{subfigure}
\usepackage{color}
\usepackage{xcolor}
\usepackage{authblk}

\newcommand{\str}[1]{\renewcommand{\baselinestretch}{#1}\normalsize}

\begin{document}

\str{1.05}

\title
{Bayesian analysis of verbal autopsy data using factor models \\ with age- and sex-dependent associations between symptoms}

\author[1]{Tsuyoshi Kunihama}
\author[2]{Zehang Richard Li}
\author[3]{Samuel J. Clark}
\author[4]{Tyler H. McCormick}
\affil[1]{Department of Economics, Kwansei Gakuin University}
\affil[2]{Department of Statistics, University of California, Santa Cruz}
\affil[3]{Department of Sociology, Ohio State University}
\affil[4]{Department of Statistics and Department of Sociology, University of Washington}

\date{March, 2024}

\maketitle

\vspace{-2mm}

\begin{abstract}

\noindent

Verbal autopsies (VAs) are extensively used to investigate the population-level distributions of deaths by cause in low-resource settings without well-organized vital statistics systems. Computer-based methods are often adopted to assign causes of death to deceased individuals based on the interview responses of their family members or caregivers. In this article, we develop a new Bayesian approach that extracts information about cause-of-death distributions from VA data considering the age- and sex-related variation in the associations between symptoms. Its performance is compared with that of existing approaches using gold-standard data from the Population Health Metrics Research Consortium. In addition, we compute the relevance of predictors to causes of death based on information-theoretic measures. \\


\end{abstract}

\str{1.55}
\section{Introduction}

Cause-of-death distributions provide fundamental information about the current dynamics of a population, which is crucial for designing, monitoring, and evaluating public health actions. However, this information is inaccessible to decision-makers in low-resource regions, where many deaths occur outside hospitals and are not officially registered. A survey-based method called verbal autopsy (VA) has been developed as a practical solution; it involves inferring the cause of death without medical certification and estimating cause-specific mortality rates in populations without complete-coverage civil registration and vital statistics systems \citep{Maher10, Sankoh12, Nkengasong20}. In VA, information about the cause of death of an individual is collected by interviewing a person close to them with questions about their demographics, medical history, and signs, and symptoms in the period leading to their death. The VA interview needs to be analyzed to assign a probable cause of death, and a common practice is to have a panel of physicians interpret the interview data. However, this consumes physicians' time for patients and is difficult to scale up for massive amounts of data. In recent years, as an alternative, algorithmic and statistical methods have been developed for cause-of-death assignment using VA data \citep{Byass03, Serina15, McCormick16}. Computer-based approaches have the advantages of being scalable for large-scale data and becoming free once their programming codes are publicly distributed. For example, popular methods are easily implemented using the openly available R package openVA \citep{Li23}. \cite{Chandramohan21} conducted a comprehensive review of the VA developments over the last several decades.

Regarding VA interview questions as predictors, statistical methods assign a cause of death based on the probability of causes given these predictors $P(\text{cause} \,|\, \text{predictors})$. Standard parametric models, such as multinomial probit/logit regression, can be applied to this conditional probability. However, VA surveys usually contain large amounts of missing values and thus require the challenging task of imputing high-dimensional data with complex interactions. The common approach is to use an indirect expression based on Bayes theorem as $P(\text{cause} \,|\, \text{predictors}) \propto P(\text{predictors}\, | \, \text{cause}) P(\text{cause})$, which makes it easy to accommodate abundant missing values under the missing-at-random assumption. Popular methods assume the conditional independence of predictors given a cause \citep{Byass12, Byass19, Miasnikof15, McCormick16}, but VA surveys can contain more than a hundred of questionnaire items and may violate this assumption. Recent studies investigated approaches to expressing complex dependent structures among predictors \citep{Li20, Kunihama20, Moran21}.

VA surveys collect various information relevant to causes of death, ranging from the demographic backgrounds to the symptoms, signs, or medical history (henceforth collectively called symptoms) of deceased individuals. All survey items are treated equally as predictors in existing methods, although some of them may play different roles in cause-of-death assignment. Demographic characteristics, especially age and sex, are key factors in many public health settings. For example, worldwide differences by sex and age have been identified in disability-adjusted life years \citep{WHO20} and the cancer burden \citep{Ferlay20}. In addition, \cite{WHO22} reported inequalities in the distribution of COVID-19 cases and deaths across different demographic profiles such that males are more likely than females to die and older people are more likely to develop serious illness. The report also shows age and sex differences in preventive behaviors during the pandemic, such as social distancing, and mask wearing. Considering the key roles of age and sex in public health, we explicitly differentiate them from other predictors and assume they affect the associations between symptoms for each cause of death. Then, assuming predictors consist of symptoms, age and sex, we consider the expression
\begin{align}
P(&\text{cause} \,|\, \text{predictors}) \propto P(\text{predictors}\, | \, \text{cause}) P(\text{cause}) \nonumber \\
&\propto P(\text{symptoms, age, sex}\, | \, \text{cause}) P(\text{cause}) \nonumber \\
&\propto P(\text{symptoms}\, | \, \text{age, sex, cause}) P(\text{age, sex} \, | \, \text{cause}) P(\text{cause}).
\label{eq:1}
\end{align}
\cite{Moran21} developed a framework where the distribution of symptoms differs by covariates, namely, $P(\text{cause} \,|\, \text{predictors}) \propto P(\text{symptoms}\, | \, \text{covariates, cause}) P(\text{cause})$. In comparison to (\ref{eq:1}), with age and sex set as covariates, the assumption of independence between (age, sex) and causes of death is implicitly imposed; that is, $P(\text{age, sex} \, | \, \text{cause}) = P(\text{age, sex})$. Given that age and sex are often associated with causes of death \citep{WHO20}, avoiding the independence assumption is appropriate.

\subsection{PHMRC VA survey data} \label{sec:data}

Using gold-standard VA data, we examine the following points: 1. The associations between symptoms vary by age and sex. 2. The distributions of age and sex differ by cause of death. We analyze neonatal VA data collected by the Population Health Metrics Research Consortium (PHMRC) project \citep{Murray11b, PHMRC13}. These data consist of medically certified causes of death and VA survey responses collected by interviewers who were blinded to the causes of death assigned in hospitals. We use five neonatal causes of death and 97 predictors for 1,619 babies, and all predictors are binarized based on the transformation by \citet{Murray11b}. Figure \ref{fig:hist} shows a histogram of the causes of death in our study, and the predictors are listed in the supplementary materials.

\begin{figure}[htbp]
  \begin{minipage}[b]{0.5\linewidth}
    \centering
    \includegraphics[keepaspectratio, scale=0.6]{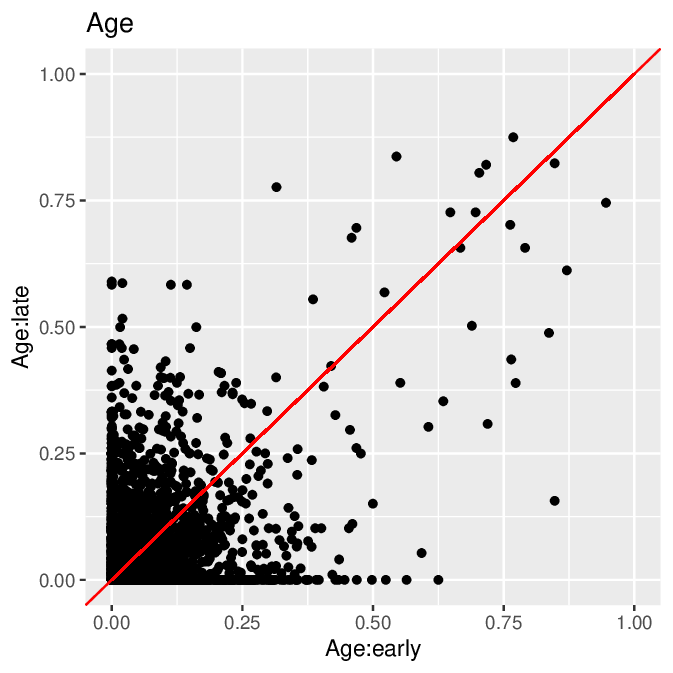}
  \end{minipage}
  \begin{minipage}[b]{0.5\linewidth}
    \centering
    \includegraphics[keepaspectratio, scale=0.6]{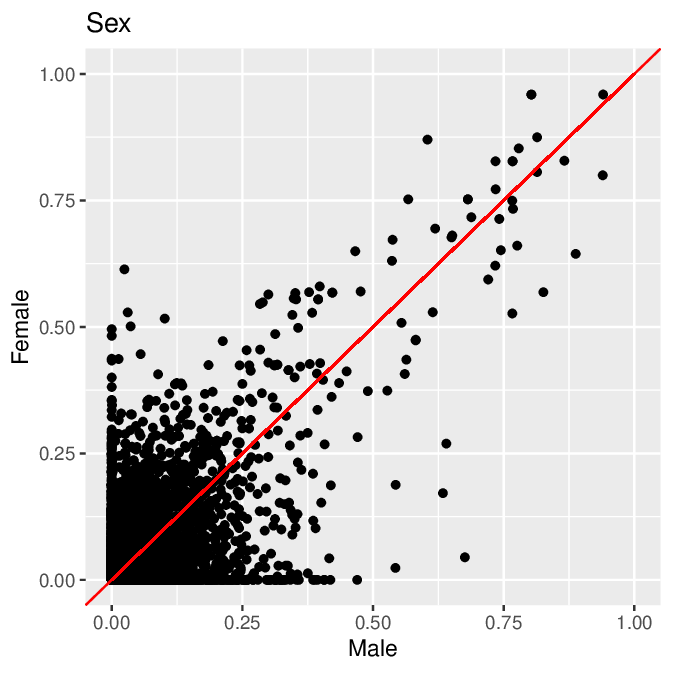}
  \end{minipage}
    \caption{Plots of Cram\'{e}r's V values for each pair of symptoms for birth asphyxia between age groups (left) and between sexes (right).}
\label{fig:1}    
\end{figure}

\begin{figure}[htbp]
  \begin{minipage}[b]{0.5\linewidth}
    \centering
    \includegraphics[keepaspectratio, scale=0.6]{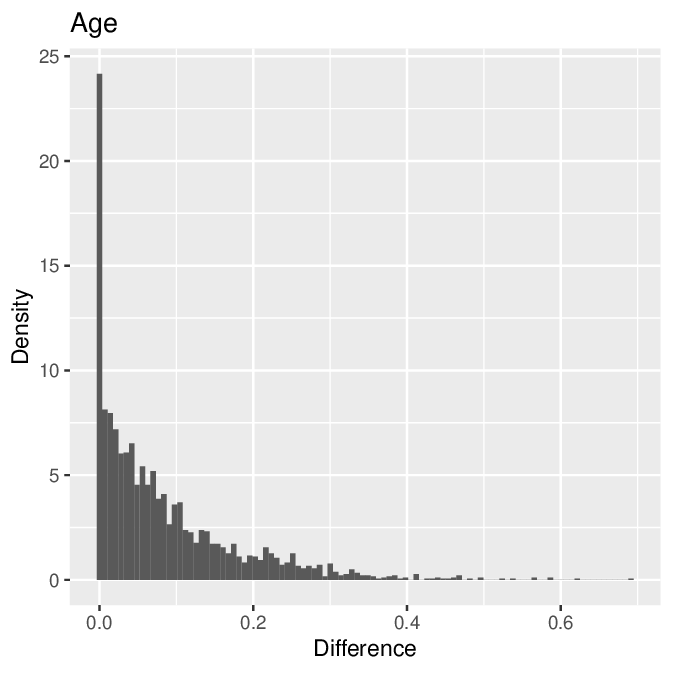}
  \end{minipage}
  \begin{minipage}[b]{0.5\linewidth}
    \centering
    \includegraphics[keepaspectratio, scale=0.6]{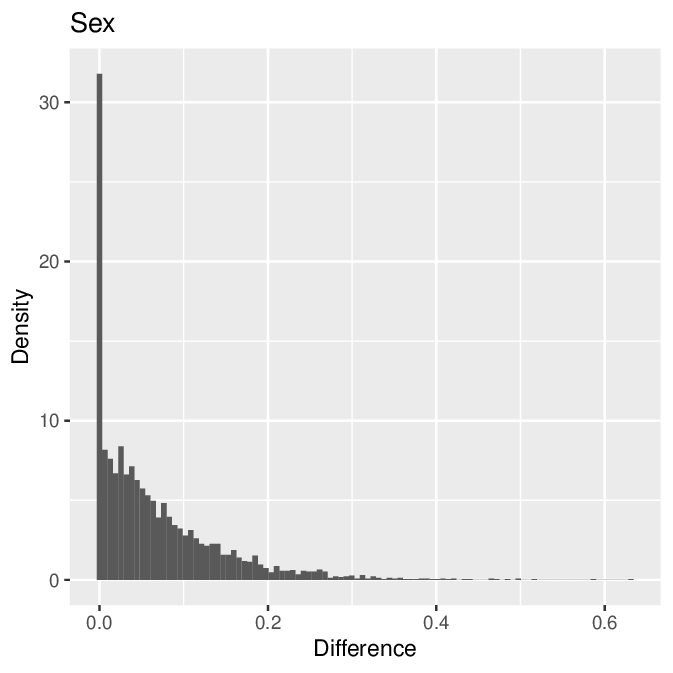}
  \end{minipage}
    \caption{Histograms of absolute difference in Cram\'{e}r's V values for each pair of symptoms for birth asphyxia between age groups (left) and between sexes (right).}
\label{fig:2}    
\end{figure}

As for the associations between symptoms, we compute Cram{\' e}r's $V$ \citep{cramer46} that measures strength of associations between two nominal variables, producing a value between 0 (no association) and 1 (complete association). Figure \ref{fig:1} shows Cram\'{e}r's V values of all pairs of symptoms for deaths by birth asphyxia between age groups and between sexes. If age and sex do not affect the associations between symptoms, then the corresponding points will be close to the diagonal lines. Figure \ref{fig:2} displays the absolute difference from each of the points to the diagonal line in Figure \ref{fig:1}, indicating that some symptom pairs show relatively large differences between age groups and between sexes. Therefore, it is worth pursuing a direction of modeling the age- and sex-dependent associations between symptoms to estimate the causes-of-death distribution.

\begin{figure}[htbp]
\centering
\includegraphics[scale=0.87]{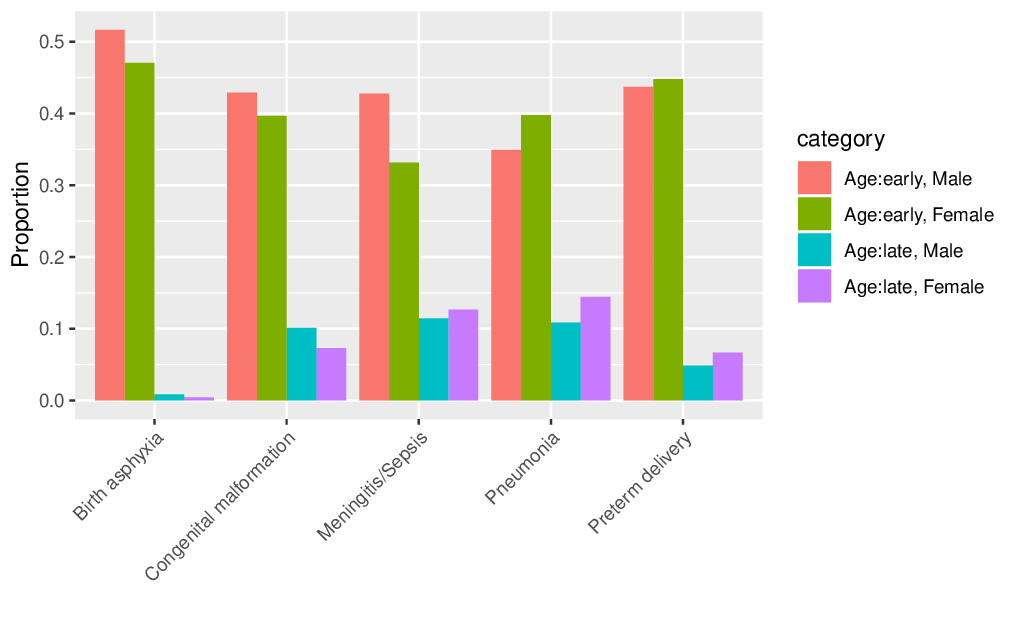}
\vspace{-5mm}
\caption{Proportions of age (early and late stages) and sex (male and female) for each cause of death in our PHMRC neonate dataset.}
\label{fig:3}
\end{figure}

Figure \ref{fig:3} shows the proportions of age groups and sexes for each cause of death. The bar patterns differ between causes. For example, most of the babies who died due to birth asphyxia are in the early age group, and the ratio of females is higher than that of males for pneumonia and preterm delivery. Hence, age and sex are considered not independent of causes of death.

The remainder of this paper is organized as follows. Section 2 shows the proposed Bayesian approach to estimating cause-of-death distributions using factor models with age- and sex-dependent associations between symptoms. Section 3 describes the Markov chain Monte Carlo (MCMC) algorithm of the proposed method. In Section 4, we assess the proposed approach in comparison to existing methods and evaluate the relevance of the predictors to the causes of death using the PHMRC neonate dataset. Section 5 concludes this article.

\begin{figure}[htbp]
\centering
\includegraphics[scale=1.1]{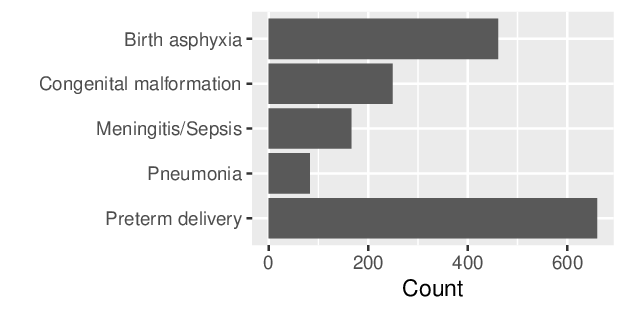}
\vspace{-5mm}
\caption{Histogram of causes of death in our PHMRC neonate dataset.}
\label{fig:hist}
\end{figure}

\section{Proposed Bayesian approach} \label{sec:models}

We develop the Bayesian models in (\ref{eq:1}) for estimating the population-level distribution of the causes of death using the VA survey data. The proposed approach divides the VA predictors into basic demographic information (age and sex) and the other questionnaire items (symptoms), and the associations between symptoms differs by age and sex. In addition, we measure the relevance of each predictor to the causes of death using the proposed models. For the notations below, let $y_i \in \{1,\ldots, L\}$ be the cause of the $i$th person's death, where $i=1,\ldots,n$, and $x_i=(x_{i1}, \ldots, x_{ip})'$ be the binary indicators of the symptoms, where $x_{ij} \in \{0, 1\}$ for $j=1,\ldots,p$. Moreover, $\text{age}_i$ and $\text{sex}_i$ indicate the age and sex, respectively, of the $i$th baby.

\subsection{Modeling of $P(\text{symptoms}\, | \, \text{age, sex, cause})$ in (\ref{eq:1})} 

First, we develop models for symptoms in which the dependence structure varies between the age and sex groups for each cause of death. Let $z_i=(z_{i1},\ldots,z_{ip})'$ be a latent continuous variable for $x_i$. The Bayesian factor (BF) model \citep{Kunihama20} is expressed as
\begin{align*}
x_{ij} &= 1( z_{ij} > 0 ), \ \ j=1,\ldots,p, \\
z_i &= \mu_{y_i} + \Lambda_{y_i} \eta_i + \varepsilon_i, \ \ \varepsilon_i \sim N( 0, I_p ), 
\end{align*}
where $1(\cdot)$ is an indicator function; $\mu_y$ and $\Lambda_y$ denote the mean and the loading matrix, respectively, for cause $y$; and $\eta_i$ is a latent factor, where $\eta_i \sim N(0,I_K)$. By integrating out the factor $\eta_i$, the dependence is induced with $\text{cov}(z_i) = \Lambda_y \Lambda'_y + I_p$. By borrowing the idea of covariance regression \citep{Hoff12, Niu19}, we place additional factors in $z_i$, leading to age- and sex-dependent associations between the symptoms.
\begin{align}
x_{ij} &= 1( z_{ij} > 0 ), \ \ j=1,\ldots,p, \label{eq:x} \\
z_i &= B_{y_i} w_i + \Lambda_{y_i} \eta_i + C_{y_i} ( w_i \otimes \gamma_i ) + \varepsilon_i, \ \ \varepsilon_i \sim N( 0, I_p ), 
\label{eq:proposed}
\end{align}
where $w_i = (1, \text{age}_i, \text{sex}_i)'$ and $C_y = [C_{y}^{(1)} \ C_y^{(2)} \ C_{y}^{(3)}]$, where $C_{y}^{(q)}$ is a $p\times G$ matrix for $q=1,2,3$; $\gamma_i$ and $\eta_i$ are latent factors, where $\eta_i \sim N(0,I_K)$ and $\gamma_i \sim N(0,I_G)$, respectively. The Kronecker product is expanded, and (\ref{eq:proposed}) is expressed as
\begin{align}
z_i &= B_{y_i} w_i + \tilde{\Lambda}_{y_i} \tilde{\eta}_i + \varepsilon_i, \ \ \varepsilon_i \sim N( 0, I_p ),
\label{eq:proposed2}
\end{align}
where $\tilde{\eta}_i = (\eta'_i, \gamma'_i)'$ is a $K+G$-dimensional latent factor, where $\tilde{\eta}_i \sim N(0, I_{K+G} )$, and $\tilde{\Lambda}_{y} = [\Lambda_y \ D_y]$ is a factor loading, where $\Lambda_y$ is constant against age and sex and $D_y$ is an age- and sex-dependent matrix given by
\begin{align*}
D_y = C_{y}^{(1)} + C_y^{(2)} \text{age}_i + C_{y}^{(3)} \text{sex}_i.
\end{align*}
For example, if $\text{age}_i\in\{0, 1\}$ and $\text{sex}_i\in\{0,1\}$ correspond to \{young, old\} and \{male, female\}, respectively, then
\begin{align*}
D_y =
\begin{cases}
C_{y}^{(1)}  &  (\text{young male}) \\
C_{y}^{(1)} + C_y^{(2)}  & (\text{old male}) \\
C_{y}^{(1)} + C_y^{(3)}  & (\text{young female}) \\
C_{y}^{(1)} + C_y^{(2)} + C_y^{(3)}  & (\text{old female}) \\
\end{cases}
\end{align*} 
By integrating $\tilde{\eta}_i$ out in (\ref{eq:proposed2}), we obtain the following covariance matrix,
\begin{align*}
\text{cov}(z_i) = \tilde{\Lambda}_y \tilde{\Lambda}'_y + I_p = \Lambda_y \Lambda'_y + D_y D'_y + I_p,
\end{align*}
where $\Lambda_y \Lambda'_y$ denotes rank-$K$ associations shared between all age and sex groups and $D_y D'_y$ denotes rank-$G$ associations changing by age and sex. To avoid complexity, we assume $K=G$ for the dimensions of the latent factors and select an appropriate value via cross-validation in our application. In general, the loading matrix need to be constrained to identify latent factors, but we use factor modeling to reduce the dimensions of the covariance matrix, in which case identification is unnecessary.

\subsection{Modeling of $P(\text{age, sex} \, | \, \text{cause})$ and $ P(\text{cause})$ in (\ref{eq:1})} 

In terms of modeling of a cause of death, age and sex, it may be standard to develop a probability of a cause given age and sex, $P(\text{cause}\,|\, \text{age}, \text{sex})$, because a cause is the variable of main interest in the VA analysis. However, our objective is not in proposing an exact data generating process of a cause given only age and sex but in developing a predictive function of a cause given all predictors. Therefore, we consider modeling $P(\text{age}, \text{sex} \,|\,\text{cause})$, which is a part of $P(\text{cause}\,|\, \text{predictors})$ derived from the expression (\ref{eq:1}).

We analyze the dichotomized version of VA data in our application and apply the Dirichlet distribution to $P(\text{age, sex} \, | \, \text{cause})$ independently for each cause.
\begin{align*}
\left\{ P(\text{age}_i= a, \text{sex}_i = b\, | \, y_i), \  a,b \in \{0,1\} \right\} \sim \text{Dirichlet}(a_1,\ldots,a_4),
\end{align*}
where $a_1,\ldots,a_4$ are concentration parameters. Likewise, regarding the prior distribution of causes $P(\text{cause})$, we assume the Dirichlet distribution
\begin{align*}
\left\{ P(y_i=l), \ l\in \{1,\ldots,L\} \right\} \sim \text{Dirichlet}(a_1,\ldots,a_L).
\end{align*}
In cases with little prior information, we assume a uniform prior with the concentration parameters that are equal to one. For example, we assume $P(y_i = l) \propto 1$ for $l=1,\ldots,L$ if little prior information is available about the distribution of causes in a new VA study site.

\subsection{Measurement of predictor relevance} \label{sec:importance}

VA surveys consist of many questionnaire items, which are conceptually useful for predicting a cause of death. However, the degree of association with causes of death can vary much from question to question, so it is of interest to investigate the relevance of each predictor to causes of death in VA survey data.

Using the proposed models in (\ref{eq:1}), we compute the strength of the associations between a cause of death and each predictor based on information-theoretic measures \citep{MacKay03, Cover06}. Let $H(y)$ denote the entropy of $y$, which quantifies the expected amount of information about $y$; $H(y) = - \sum_y P(y) \log P(y)$. Mutual information (MI) is a measure of the mutual dependence between two random variables. Letting $\tilde{x}=(\text{age},\text{sex},x)$, MI is defined by
\begin{align}
I(y;\tilde{x}_j) = \sum_{y} \sum_{\tilde{x}_j} P(y, \tilde{x}_j) \log \frac{ P(y, \tilde{x}_j) }{P(y) P(\tilde{x}_j)}.
\label{eq:def-mi}
\end{align}
This denotes the amount of information obtained about a cause $y$ by observing the $j$th predictor $\tilde{x}_j$. Theoretically, (\ref{eq:def-mi}) takes a nonnegative value and equals zero if and only if the two random variables are independent. To improve our interpretation, we normalize the MI to take a value in the $[0,1]$ interval, we normalize MI by dividing it by the entropy of $y$; that is $I(y;\tilde{x}_j)/H(y)$, which indicates the amount of information about causes added by the $j$th predictor relative to all the information about causes.

VA questionnaires often have multiple questions on each topic under symptoms. For example, the PHMRC survey on neonatal death asks six questions each about suckling and breathing. If these questions provide highly overlapped information about causes, then some of them may show high MI values despite adding little new information about causes given the other questions. Considering this point, we compute the conditional MI (CMI) along with the MI. The CMI indicates the amount of information each predictor can add given the other items.
\begin{align}
I(y;\tilde{x}_j\,|\,\tilde{x}_{-j}) = \sum_{y} \sum_{\tilde{x}} P(y, \tilde{x}) \log \frac{ P(y, \tilde{x}_j\,|\,\tilde{x}_{-j}) }{P(y\,|\,\tilde{x}_{-j}) P(\tilde{x}_j\,|\,\tilde{x}_{-j})},
\label{eq:def-cmi}
\end{align}
where $\tilde{x}_{-j}$ indicates all the predictors except the $j$th one. We normalize the CMI to $I(y;\tilde{x}_j\,|\,\tilde{x}_{-j})/H(y)$, which assumes a value in the $[0,1]$ interval and indicates the amount of additional information about causes we can obtain by knowing $x_j$ given all the other predictors relative to the total amount of information related to causes. The CMI can capture the strength of the conditional association between the predictors and causes of death, but relying only on CMI may result in a misleading conclusion. For example, if two predictors are strongly associated with causes but share most of their information, then both of them will show small CMI values despite being useful for predicting the causes of death. Therefore, the relevance of predictors should be evaluated by investigating both the MI and CMI values. The computation of the MI and CMI using the proposed models is detailed in Section \ref{sec:posterior}.

\section{Posterior computation}  \label{sec:posterior}

We develop an Markov-chain Monte Carlo (MCMC) algorithm for the proposed method in Section \ref{sec:models}. Let $m_i = (m_{i1},\ldots,m_{ip})'$ be a vector of indicators denoting missing values for the $i$th person such that $m_{ij}=1$ if $x_{ij}$ is missing and $m_{ij}=0$ if $x_{ij}$ is observed, where $j=1,\ldots,p$.  We define the notation $[m_i]$ such that $a_{[m_i]}$ and $A_{[m_i]}$ (where $a$ and $A$ are a vector and a matrix with $p$ rows, respectively) denote a subvector and a submatrix consisting of components with $m_{ij}=0$ for $j=1,\ldots,p$. In addition, $A_{j\cdot}$ shows the $j$th row of the matrix $A$. 

For the prior distributions on $B_y$, $\Lambda_y$ and $C_y$, we use Cauchy distributions. It has high density around zero and heavy tails, which reduce the effects of redundant elements while capturing meaningful signals. Based on the expression of the Cauchy distribution via normal-gamma distributions, we assume $B_{yjq} \sim N(0, \phi^{-1}_{Bjq})$, $\phi_{Bjq} \sim Ga(0.5, 0.5)$, $\Lambda_{yjk} \sim N(0, \phi^{-1}_{\Lambda j})$, $\phi_{\Lambda j} \sim Ga(0.5, 0.5)$, $C^{(q)}_{yjk}\sim N(0, \phi^{-1}_{Cjq})$, $\phi_{Cjq} \sim Ga(0.5, 0.5)$ where $q=1,2,3$, $j=1,\ldots,p$, $k=1,\ldots,K$, and $Ga(a,b)$ denotes a gamma distribution with a mean $a/b$. The latent variables $\phi_{B jq}$, $\phi_{\Lambda j}$ and $\phi_{C jq}$ are shared between causes and factors to reduce the number of model parameters.

The proposed MCMC algorithm is below.

\begin{enumerate}

\item Update $\beta_{y j} \equiv ( B_{yj \cdot}, \Lambda_{yj\cdot}, C^{(0)}_{yj\cdot}, C^{(1)}_{yj\cdot}, C^{(2)}_{yj\cdot})'$ from $N(\mu_*, \Sigma_*)$ for $y=1,\ldots,L$ and $j=1,\ldots,p$ with
\begin{align*}
\mu_* = \Sigma_* \left( \sum_{i:y_i=y, m_{ij}=0} a_i z_{ij} \right), \ \ \Sigma_* = \left( \sum_{i:y_i=y, m_{ij}=0} a_i a'_i + \Sigma_0^{-1} \right)^{-1}
\end{align*} 
where $a_i = (w'_i, \eta'_i, \gamma'_i, \text{age}_i \gamma'_i , \text{sex}_i\gamma'_i)'$ and \\ $\Sigma_0^{-1} = \text{diag}(\phi_{Bj1}, \phi_{Bj2}, \phi_{Bj3}, \phi_{\Lambda j} 1_K, \phi_{Cj1}1_K, \phi_{Cj2}1_K, \phi_{Cj3}1_K)$, where $1_K$ denote a $K$-dimensional vector with all elements one.

\item Update $\tilde{\eta}_i = (\eta'_i, \gamma'_i)'$ from $N(\tilde{\mu}, \tilde{\Sigma})$ for $i=1,\ldots,n$ with
\begin{align*}
\tilde{\mu} = \tilde{\Sigma} \tilde{\Lambda}'_{y_i [m_i]} (z_i - B_{y_i}w_i )_{[m_i]}, \ \ \tilde{\Sigma} = \left( \tilde{\Lambda}'_{y_i [m_i]}  \tilde{\Lambda}_{y_i [m_i]} + I_{2K} \right)^{-1}.
\end{align*} 

\item Update $\phi_{Bjq}$ from $Ga ( 0.5 (L+ 1), 0.5 ( \sum_{y=1}^L B^2_{yjq} + 1) )$ for $q=1,2,3$ and $j=1,\ldots,p$.

\item Update $\phi_{\Lambda j}$ from $Ga( 0.5(LK+ 1), 0.5 \sum_{y=1}^L \sum_{k=1}^K  \Lambda^2_{y jk} + 1) )$ for $j=1,\ldots,p$.

\item Update $\phi_{C jq}$ from $Ga( 0.5(LK+ 1), 0.5 \sum_{y=1}^L \sum_{k=1}^K  C^{(q)2}_{y jk} + 1) )$ for $q=1,2,3$ and $j=1,\ldots,p$.

\item Update $z_{ij}$ with $m_{ij}=0$ for $i=1,\ldots,n$ and $j=1,\ldots,p$ from 
\begin{align*}
\begin{cases}
N_+( B_{y_i j \cdot} w_i + \tilde{\Lambda}_{y_i j \cdot} \tilde{\eta}_i, 1) & \text{if $x_{ij} =1$}, \\
N_-( B_{y_i j \cdot} w_i + \tilde{\Lambda}_{y_i j \cdot} \tilde{\eta}_i, 1) & \text{if $x_{ij} = 0$}, \\
\end{cases}
\end{align*}
where $N_+$ and $N_-$ denote truncated normals with support $[0, \infty)$ and $(-\infty, 0]$.

\item Update $\left\{ P(\text{age}_i= a, \text{sex}_i = b\, | \, y_i=y), \  a,b \in \{0,1\} \right\} $ from $\text{Dirichlet}(b_{00},b_{01},b_{10},b_{11})$ where \\ $b_{ab} = 1 + \sum_{i=1}^n 1(y_i = y, \ \text{age}_i = a, \ \text{sex}_i = b)$ for $y=1,\ldots,L$.

\item Impute $\text{age}_i$ and $\text{sex}_i$ if missing from 
\begin{align*}
P(\text{age}_i = a \,|\, y_i, x_i, \text{sex}_i, \tilde{\eta}_i) &= \frac{ P(x_i\,|\,y_i,\text{age}_i=a,\text{sex}_i, \tilde{\eta}_i) P(\text{age}_i=a, \text{sex}_i \,|\, y_i)  }{ \sum_{l=0}^1 P(x_i\,|\,y_i,\text{age}_i=l,\text{sex}_i, \tilde{\eta}_i) P(\text{age}_i=l, \text{sex}_i \,|\, y_i) }, \ \ a=0,1, \\
P(\text{sex}_i = b \,|\, y_i, x_i, \text{age}_i, \tilde{\eta}_i) &= \frac{ P(x_i\,|\,y_i,\text{age}_i,\text{sex}_i=b, \tilde{\eta}_i) P(\text{age}_i, \text{sex}_i=b \,|\, y_i)  }{ \sum_{l=0}^1 P(x_i\,|\,y_i,\text{age}_i,\text{sex}_i=l, \tilde{\eta}_i) P(\text{age}_i, \text{sex}_i=l \,|\, y_i) }, \ \ b=0,1.
\end{align*}
where $P(x_i\,|\,y_i,\text{age}_i,\text{sex}_i, \tilde{\eta}_i)$ corresponds to the probability from (\ref{eq:x}) and (\ref{eq:proposed2}).

\item For $i \in S$ where $S$ is target data, predict $y_i$ with
\begin{align*}
P(y_i = y \,|\, x_i, \text{age}_i, \text{sex}_i) = \frac{ P(x_i \,|\, y_i=y, \text{age}_i, \text{sex}_i) P(\text{age}_i, \text{sex}_i\,|\,y_i=y) P(y_i=y) }{ \sum_{l=1}^L P(x_i \,|\, y_i=l, \text{age}_i, \text{sex}_i) P(\text{age}_i, \text{sex}_i\,|\,y_i=l) P(y_i=l) }, 
\end{align*}
with $y=1,\ldots, L$ where $P(x_i \,|\, y_i, \text{age}_i, \text{sex}_i) = \int P(x_i \,|\, y_i, \text{age}_i, \text{sex}_i, \tilde{\eta}) f(\tilde{\eta}) d\tilde{\eta}$ is evaluated via a Monte Carlo approximation with $\tilde{\eta}_{(r)} \sim N(0, I_{2K})$ for $r=1,\ldots,R$,
\begin{align}
P(x_i \,|\, y_i, \text{age}_i, \text{sex}_i)  \approx \frac{1}{R} \sum_{r=1}^R \left\{ \prod_{j:m_{ij}=0} P(x_{ij} \,|\, \tilde{\eta}_{(r)}, y_i,, \text{age}_i, \text{sex}_i) \right\}.
\label{eq:mcmc-r}
\end{align}
Then, compute the population distribution of the causes of death as
\begin{align*}
\left( \frac{1}{ n_S } \sum_{i \in S} 1( y_i = 1 ), \ldots, \frac{1}{ n_S } \sum_{i \in S} 1( y_i = L ) \right)
\end{align*}
where $n_S$ is the number of observations in the test data. For the computation of the measures of importance of the predictors, Step 9 is replaced as follows:
\item[9.] Update $\left\{ P (y_i = y), \ y \in \{1,\ldots,L\} \right\}$ from
\begin{align*}
\text{Dirichlet}\left( 1 + \frac{1}{ n } \sum_{i=1}^n 1( y_i = 1 ), \ldots, 1+\frac{1}{ n } \sum_{i =1}^n 1( y_i = L ) \right).
\end{align*}
\item[10.] Compute the MI and CMI for each predictor in (\ref{eq:def-mi}) and (\ref{eq:def-cmi}). Because the summation in (\ref{eq:def-cmi}) is intractable for a nonsmall $p$, it is approximated using the Monte Carlo sample $(y_{(r)},\tilde{x}_{(r)}) \sim P(y,\tilde{x}) = P(x\,|\,y,\text{age},\text{sex})P(\text{age},\text{sex}\,|\,y) P(y)$, where $r=1,\ldots,\tilde{R}$.
\begin{align}
I(y;\tilde{x}_j\,|\,\tilde{x}_{-j}) \approx \frac{1}{\tilde{R}} \sum_{r=1}^{\tilde{R}} \log \frac{ P(y_{(r)}, \tilde{x}_{(r)j}\,|\,\tilde{x}_{(r)-j}) }{P(y_{(r)}\,|\,\tilde{x}_{(r)-j}) P(\tilde{x}_{(r)j}\,|\,\tilde{x}_{(r)-j})}.
\label{eq:cmi-approx}
\end{align}

\end{enumerate}
The programming codes for the proposed method are distributed on GitHub (https://github.\\com/kunihama/BF-AS).

\section{Results} \label{sec:result}

In this section, we evaluate the performance of the proposed approach using the PHMRC neonatal VA data. Our primary interest is in estimating the distribution of deaths by cause in a target population. In VA studies, it is often extremely challenging to obtain gold-standard training data from a target site where the VA survey questions are tied to the medically certified cause of death of each individual. Therefore, the cause-of-death distribution in a target site is predicted using training data collected from different geographic locations, so the obtained fractions of death by cause vary widely between testing and training data. Considering this point, we consider realistic scenarios where each PHMRC study site is treated as a target site and all other sites are jointly regarded as training data.
Thus, we have six cases where the targets are Andhra Pradesh, India (case 1), Bohol, Philippines (case 2), Dar es Salaam, Tanzania (case 3), Mexico City, Mexico (case 4), Pemba Island, Tanzania (case 5) and Uttar Pradesh, India (case 6). 

We assess VA classification methods based on the accuracy of their cause-specific mortality fractions (CSMFs) \citep{Murray11c}.
\begin{align*}
\text{CSMF accuracy} = 1 - \frac{ \sum_{y=1}^L \left| P_0(y_i=y) - P(y_i=y) \right| }{ 2\{ 1 - \min_{1 \leq y \leq L} P_0(y_i=y) \} }
\end{align*}
where $P_0$ and $P$ are the true and estimated distributions, respectively, of the causes of death in a target site. The CSMF measures the closeness between the two distributions, taking a value in the [0,1] interval, and a larger value means better performance. In our study, the true distribution $P_0$, is approximated using the empirical distribution of causes in the test data, whereas $P$ corresponds to the distribution of causes estimated using a classification tool. 

The proposed method (BF models with age and sex dependence [BF-AS]) is compared with recent VA classification tools. Popular methods in the VA literature are available in the R package, such as openVA \citep{Li23} with InterVA \citep{Byass12}, the tariff method \citep{James11, Serina15}, the naive Bayes classifier (NBC) \citep{Miasnikof15} and InSilicoVA \citep{McCormick16}. These approaches are based on the conditional independence of symptoms given a cause, whereas the BF models of \cite{Kunihama20} and the Bayesian hierarchical factor regression (FARVA) model of \cite{Moran21} describe the dependence between symptoms for each cause. FARVA allows the associations between symptoms to vary by covariate, which we set as age and sex. The settings of the competitor methods are detailed in the supplementary materials. For the proposed method, we standardized $\text{age}_i$ and $\text{sex}_i$ in $w_i$ in (\ref{eq:proposed}) with the mean zero and the standard deviation one. And we selected the number of latent factors $K$ from the set $\{1,2,3,4,5\}$ via fivefold cross-validation and set $R=1,000$ in (\ref{eq:mcmc-r}) and $\tilde{R}=10,000$ in (\ref{eq:cmi-approx}) as the number of random samples for the Monte Carlo approximation. We generated 10,000 MCMC samples after the initial 1,000 burn-in iterations, and every 20th sample was saved. We observed that the sample paths were stable, and the sample autocorrelations dropped smoothly. The trace plot and the autocorrelation plot of the CSMF in the target site are in the supplementary materials.

\begin{figure}[htbp]
\centering
\includegraphics[scale=0.85]{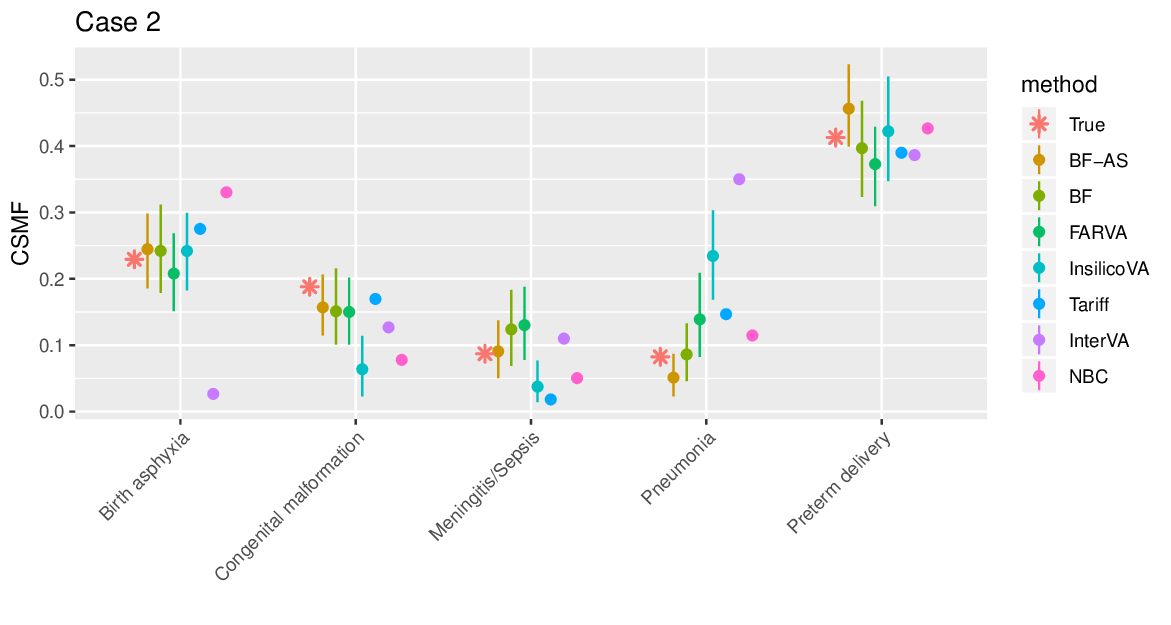}
\vspace{-5mm}
\caption{CSMF estimation results for case 2. The red asterisks show the true values, and each circle with an interval corresponds to the mean and the 95\% interval, respectively, of each statistical method. BF-AS: BF models with age and sex dependence (proposed method), BF: Bayesian factor models, FARVA: Bayesian hierarchical factor regression model, NBC: naive Bayes classifier. Tariff, InterVA, and NBC produce point estimates only.}
\label{fig:case2}
\end{figure}

\begin{figure}[htbp]
\centering
\includegraphics[scale=0.7]{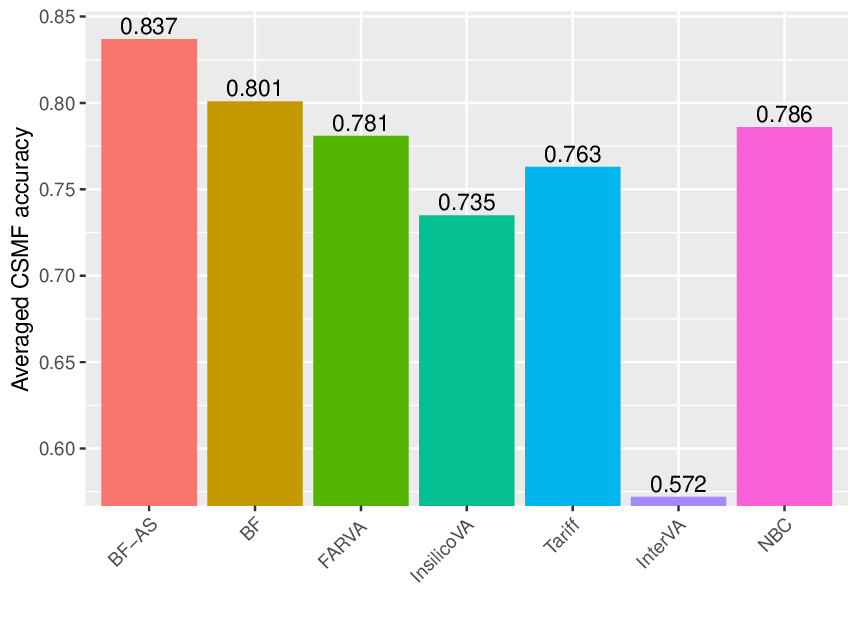}
\vspace{-5mm}
\caption{Average CSMF accuracy of each method in six cases.}
\label{fig:comparison}
\end{figure}

First, as an illustrative example, Figure \ref{fig:case2} shows the CSMF estimation results for case 2. All methods except InterVA work reasonably well, capturing the true mortality fractions closely; InterVA underestimates birth asphyxia while overestimating pneumonia. The figures for the other cases are in the supplementary materials. Table \ref{table:comparison} reports the CSMF accuracy of all methods, with the bold numbers indicating the top three results in each case. No single method dominates its competitors all the time; the approach with the highest accuracy differs from case to case. However, as the bold numbers show, the proposed method consistently performs well in all cases. In addition, Figure \ref{fig:comparison} shows the average CSMF accuracy of each method over the six cases. InterVA lags by a large margin, and although the BF models produce the highest accuracy among the competing methods, BF-AS outperforms them.

Regarding uncertainty, the coverage rate of the true fractions by the proposed method in a 95\% interval is 0.6. Although it is larger than those of the BF models (0.43), FARVA (0.4), and InSilicoVA (0.3) (Tariff, InterVA, and NBC produce only point estimates), certain gaps need to be addressed in future work, as discussed in Section \ref{sec:conclusion}.
 
\begin{table}[t]
  \centering
  \begin{tabular}{l|cccccc}
    \hline
    Method  & Case 1 & Case 2 & Case 3 & Case 4 & Case 5 & Case 6  \\
    \hline
    BF-AS 	& {\bf 0.757} & {\bf 0.932} & {\bf 0.769} & {\bf 0.893} & {\bf 0.845} & {\bf 0.822} \\
    BF 	& 0.741 & {\bf 0.942} & 0.742 & 0.788 & {\bf 0.792} & 0.801 \\
    FARVA 	& 0.734  & {\bf 0.891} & {\bf 0.768} & {\bf 0.806} & 0.740 & 0.746 \\
    InsilicoVA & 0.671 & 0.810 & 0.706 & 0.707 & 0.628 & {\bf 0.884} \\
    Tariff 	& {\bf 0.777}  & 0.880 & {\bf 0.811} & 0.566 & {\bf 0.800} & 0.740 \\
    InterVA 	& 0.471 & 0.683 & 0.463 & 0.612 & 0.659 & 0.539 \\    
    NBC 	& {\bf 0.759} & 0.840 & 0.716 & {\bf 0.837} & 0.685 & {\bf 0.877} \\
    \hline
  \end{tabular}
  \caption{CSMF accuracy. The bold numbers indicate the top three results in each case. BF-AS: BF models with age and sex dependence (proposed method), BF: Bayesian factor models, FARVA: Bayesian hierarchical factor regression model, NBC: naive Bayes classifier. The posterior means are used as the estimate of the CSMF for BF-AS, BF, FARVA and InsilicoVA.}
\label{table:comparison}  
\end{table}

\begin{figure}[htbp]
\centering
\includegraphics[scale=1]{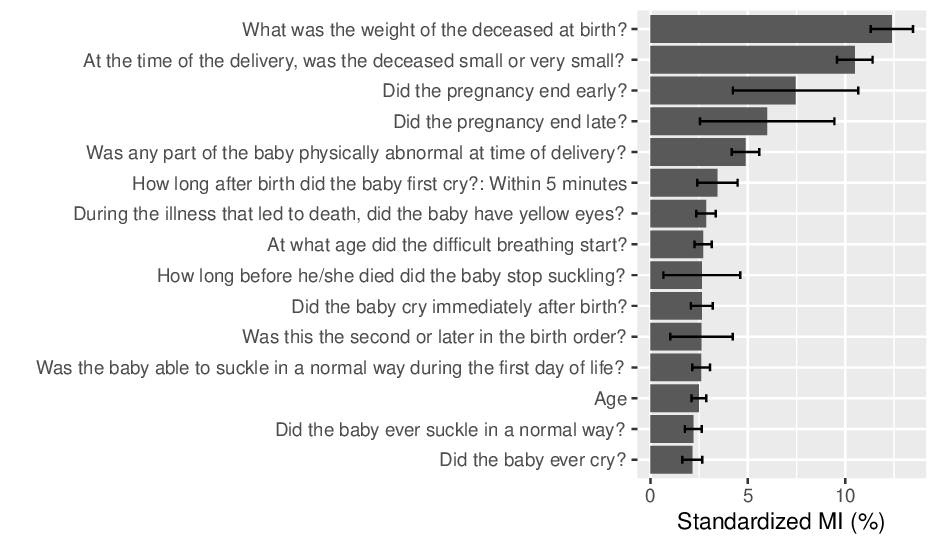}
\vspace{-5mm}
\caption{Posterior mean of standardized MI for top 15 predictors. The error bars correspond to standard deviations.}
\label{fig:mi}
\end{figure}

About the relevance of the predictors, Figure \ref{fig:mi} shows the estimated standardized MI for the top 15 predictors. The first two predictors, weight and size at the time of delivery, have more than 10\% of the information about the causes of death. Early/late delivery, physical abnormality at birth and several items related to crying and suckling also have high MI values. As for the demographic variables, age is on the list, but sex is out.

\begin{figure}[htbp]
\centering
\includegraphics[scale=1]{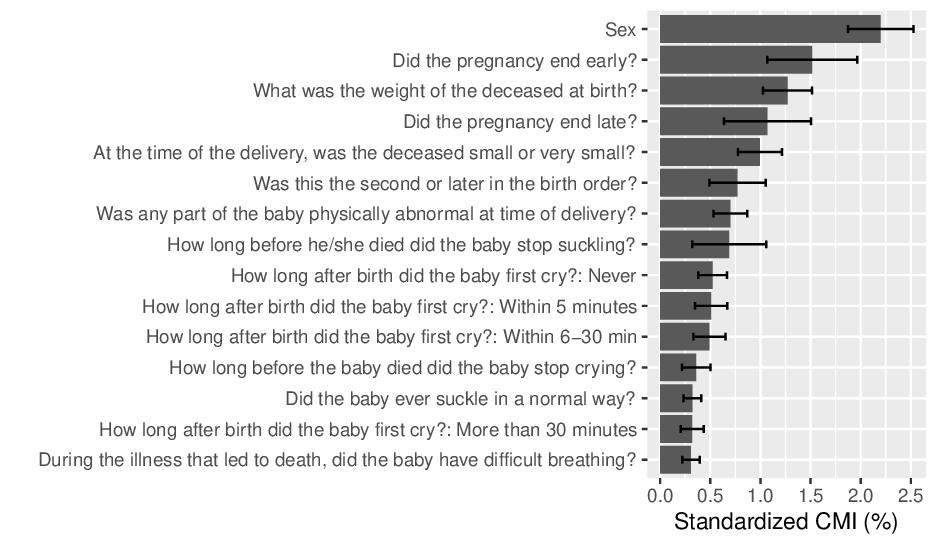}
\vspace{-5mm}
\caption{Posterior mean of standardized CMI for top 15 predictors. The error bars correspond to standard deviations.}
\label{fig:cmi}
\end{figure}

As discussed in Section \ref{sec:importance}, the MI of a predictor does not reflect the effects of other predictors. Figure \ref{fig:cmi} reports the estimated CMI for the top 15 predictors. Many of the predictors with large MI values appear again on the CMI list of CMI, such as weight, size, early/late delivery, physical abnormality, birth order and symptoms related to crying and suckling. The estimated values of CMI values for these predictors are smaller than the estimated MI values, indicating that a certain amount of their information about the causes of death is shared with the other predictors. In addition, we observe a remarkable difference --the position of sex-- between Figures \ref{fig:mi} and \ref{fig:cmi}. Sex shows a small posterior mean of the MI (0.15\%), ranking 96th out of the 97 predictors (not shown in this figure), but leads the CMI list by a wide margin. It is also the only predictor whose CMI value is larger than its MI value. Thus, although one does not obtain so much information about causes only by knowing the sex of a baby, it can indirectly deliver unique information about the cause of death through the other predictors.

\begin{figure}[htbp]
\centering
\includegraphics[scale=0.82]{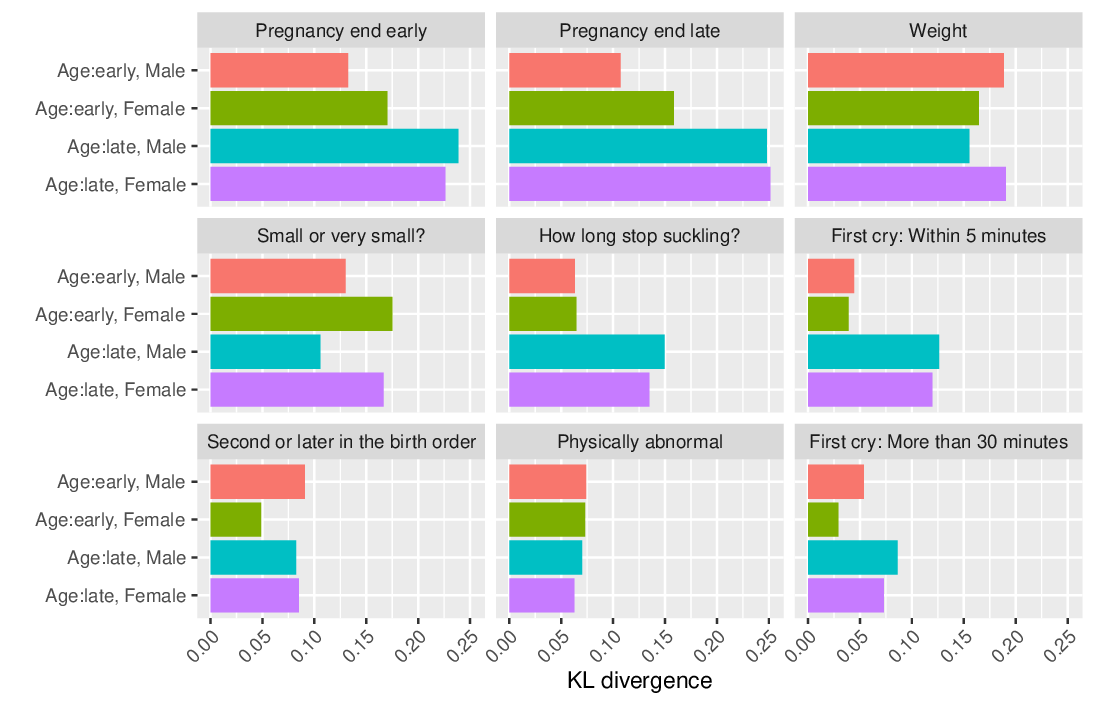}
\vspace{-5mm}
\caption{Posterior means of the KL divergence for cause of death and symptom associations across age and sex groups.}
\label{fig:KL}
\end{figure}

Finally, we investigate the variation in the association between a cause and each symptom across different age and sex groups. The mutual information, conditional on age and sex, is expressed as the expected value of the Kullback–Leibler (KL) divergence between the joint distribution and the product of marginal distributions,
\begin{align*}
&I(\text{cause}; \text{symptom} \,|\, \text{age}, \text{sex}) \\ &= E_{\text{age}, \text{sex}}\left[ D_{\text{KL}} \{P(\text{cause}, \text{symptom} \,|\, \text{age}, \text{sex}) \, || \, P(\text{cause} \,|\, \text{age}, \text{sex}) P(\text{symptom} \,|\, \text{age}, \text{sex}) \}  \right], \\
&= \sum_{\text{age}} \sum_{\text{sex}} P(\text{cause}, \text{symptom}, \text{age}, \text{sex}) \log \frac{P(\text{cause}, \text{symptom} \,|\, \text{age}, \text{sex}) }{ P(\text{cause} \,|\, \text{age}, \text{sex}) P(\text{symptom} \,|\, \text{age}, \text{sex})}. 
\end{align*}
To examine the differences between age and sex groups, Figure \ref{fig:KL} presents a comparison of the KL divergence for symptoms that are highly relevant to causes of death, revealing diverse patterns. For instance, the KL divergence indicates larger values for older age groups (pregnant end early/late, stop sucking and first cry within 5 minutes) and for females (small or very small). Conversely, there is less variation in the KL divergence by age and sex concerning the physically abnormal symptom. Additionally, females in the early age stage are associated with relatively smaller KL divergence values (birth order and first cry more than 30 minutes). These findings suggest that the degree of association between a cause of death and each symptom varies significantly by age and sex.

\section{Conclusion} \label{sec:conclusion}

In this article, we develop an approach to estimating cause-of-death distributions using VA survey data. Considering the differences between basic demographic information (age and sex) and the other types of predictors, we propose BF models with age- and sex-dependent associations between symptoms for each cause of death. In addition, the relevance of predictors to causes of death is computed based on information-theoretic measures. In our analysis of PHMRC neonate data, BF-AS performs well in estimating the cause-of-death distributions in the target sites compared with existing methods in the VA literature. However, as the proposed approach has low coverage rates of the true fractions of causes in the 95\% interval, it needs further development for a better understanding of the relations between causes of death and VA survey items.

One research direction is to allow cause-predictor relations to differ flexibly by location. In the proposed method, the distribution of predictors given a cause is independent of the VA survey site, but a recent work stated that this conditional probability can be affected by the domain \citep{Wu21, Li21}. One approach may be to treat the domain information similarly to age and sex in the proposed model by adding site-related dummy variables as covariates. However, this is challenging because target sites often do not have training data, so domain effects on symptom distributions can not be estimated directly. Domains may have similar patterns of the relations between causes and VA questions, so borrowing information from other sites for incorporation into the factor models is key.

Another study direction is to investigate whether the estimation of cause-of-death distributions will be improved by allowing symptom associations to vary by other demographic variables along with age and sex. For example, educational backgrounds, income levels and marital status may affect the symptom distribution for each cause. Adding information will increase the flexibility of factor models, but sparsity may have to be incorporated into the models to avoid overfitting, such as by using variable selection methods for covariance regression \citep{Niu19}.

\section*{Acknowledgment} 

This work was supported by JSPS KAKENHI Grant Numbers JP21K13274, JP20KK0292, JP19H00588. The results are generated mainly using Ox \citep{Doornik07}. The programing codes for the proposed method and the Bayesian factor model by \cite{Kunihama20} are distributed on the GitHub website (https://github.com/kunihama/BF-AS).

\bibliographystyle{chicago}
\bibliography{VA}

\end{document}


\title
{Supplementary materials for ``Bayesian analysis of verbal autopsy data using factor models with age- and sex-dependent associations between symptoms''}

\date{}

\maketitle

\vspace{-8mm}

\section{Lists of predictors in our analysis}

Tables 1-3 show the lists of 97 predictors in our analysis ($n=1,619$). We selected these questions from sections 1, 2 and 3 in the VA questionnaire for neonate and child in \cite{PHMRC13}, removing overlapped items and very rare symptoms only less than 10 babies exhibited. Following the approach by \cite{Murray11b}, the non-binary predictors are transformed into dichotomous variables. 

\section{Setting of the competitors}

We estimate InterVA, Tariff, the naive Bayes classifier and InSilicoVA using the R package openVA \citep{Li23} with the default setting discussed in \cite{McCormick16}. In addition, we estimate the Bayesian hierarchical factor regression (FARVA) with age- and sex-dependent associations between symptoms using the farva package (https://github.com/kelrenmor/farva). Following the settings in the example file in the package, the number of factors was set as 10 and we ran the MCMC for $10,000$ iterations and saved every 10th iteration after discarding the first half of the chain. 

\section{MCMC convergence}

To investigate posterior convergence in our applications in Section 4, the sample paths and the autocorrelations of the MCMC sample of the CSMF in the target site by the proposed model are shown in Figures 4 (case 1 and case 2), 5 (case 3 and case 4) and 6 (case 5 and case 6).

\section{CSMF estimation results}

Figures 1, 2 and 3 show the CSMF estimation results for cases 1 and 2 (Figure 1), 3 and 4 (Figure 2) and 5 and 6 (Figure 3), respectively. 

\begin{table}[h]
\centering
\caption{List of predictors: 1-40}
\begin{tabular}{cl}
\hline
\multicolumn{1}{c}{No.}		& \multicolumn{1}{l}{Predictors}	 \\
\hline 
1 & Age (How old was the deceased at the time of death?) \\
2 & Sex  \\
3 & Was the deceased a singleton or multiple birth? \\ 
4 & Was this the first, second, or later in the birth order? \\ 
5 & Is the mother still alive? \\ 
6 & Was the deceased not born in a health facility? \\ 
7 & At the time of the delivery, was the deceased small or very small? \\ 
8 & What was the weight of the deceased at birth? \\ 
9 & Was the child born alive or dead? \\ 
10 & Did the baby ever cry? \\
11 & Did the baby ever move? \\ 
12 & Did the baby ever breathe? \\ 
13 & For questions 10, 11 and 12, all three responses are No? \\ 
14 & How old was the baby/child when the fatal illness started? \\ 
15 & How long did the illness last? \\
16 & Did the deceased die at home or on route to a health facility? \\ 
17 & Was the late part of the pregnancy, labor or delivery complicated by convulsions? \\ 
18 & $\ldots$ complicated by high blood pressure? \\ 
19 & $\ldots$ complicated by severe anemia? \\ 
20 & $\ldots$ complicated by diabetes? \\ 
21 & $\ldots$ complicated by child delivered not head first? \\ 
22 & $\ldots$ complicated by cord delivered first? \\ 
23 & $\ldots$ complicated by cord around child's neck? \\
24 & $\ldots$ complicated by excessive bleeding? \\ 
25 & $\ldots$ complicated by fever during labor? \\ 
26 & Did the pregnancy end early? \\ 
27 & Did the pregnancy end late? \\
28 & Was the baby moving in the last few days before the birth? \\ 
29 & When did the mother last feel the baby move? \\ 
30 & Did the water break before labor or during labor? \\ 
31 & How much time before labor did the water break? \\ 
32 & Was the color of the liquid not clear when it broke? \\ 
33 & Was the liquor foul smelling? \\ 
34 & How much time did the labor and delivery take? \\ 
35 & Did the mother receive any vaccinations since reaching adulthood including \\ & during this pregnancy? \\
36 & Did the mother receive fewer than 3 doses of vaccine since reaching adulthood? \\ 
37 & Did the delivery not occur in a health facility? \\ 
38 & Was the delivery carried out by someone other than a health professional? \\ 
39 & Was the delivery...? Vaginal with Forceps \\ 
40 & Was the delivery...? Vaginal without Forceps \\ 
\hline
\end{tabular}
\end{table}

\begin{table}[h]
\centering
\caption{List of predictors: 41-80}
\begin{tabular}{cl}
\hline
\multicolumn{1}{c}{No.}		& \multicolumn{1}{l}{Predictors}	 \\
\hline 
41 & Was the delivery…? Vaginal Don't Know \\ 
42 & Was the delivery…? C-Section \\ 
43 & During labor but before delivery, did the mother receive any kind of injection? \\
44 & Were there any bruises or signs of injury on the baby’s body at birth? \\
45 & Was any part of the baby physically abnormal at time of delivery? \\
46 & What were the abnormalities? Head size very small at time of birth \\
47 & What were the abnormalities? Head size very large at time of birth? \\ 
48 & What were the abnormalities? Mass defect on the back of head or spine \\ 
49 & What were the abnormalities? Other \\ 
50 & Did the baby breathe immediately after birth? \\ 
51 & Did the baby have difficulty breathing? \\ 
52 & Was anything done to try to help the baby breathe at birth? \\ 
53 & Did the baby cry immediately after birth? \\ 
54 & How long after birth did the baby first cry? Within 5 minutes \\
55 & How long after birth did the baby first cry? Within 6-30 minutes \\ 
56 & How long after birth did the baby first cry? More than 30 minutes \\ 
57 & How long after birth did the baby first cry? Never \\ 
58 & Did the baby stop being able to cry? \\ 
59 & How long before the baby died did the baby stop crying? One day or more? \\ 
60 & Was the baby able to suckle in a normal way during the first day of life? \\ 
61 & Did the baby ever suckle in a normal way? \\ 
62 & Did the baby stop being able to suckle in a normal way? \\ 
63 & How long after birth did the baby stop suckling? \\ 
64 & How long before he/she died did the baby stop suckling? One day or more? \\ 
65 & Was the baby able to open his/her mouth at the time he/she stopped sucking? \\ 
66 & During the illness that led to death, did the baby have difficult breathing? \\ 
67 & At what age did the difficult breathing start? \\
68 & For how many days did the difficult breathing last? \\ 
69 & During the illness that led to death, did the baby have fast breathing? \\ 
70 & At what age did the fast breathing start? \\ 
71 & For how many days did the fast breathing last? \\
72 & During the illness that led to death, did the baby have indrawing of the chest? \\ 
73 & During the illness that led to death, did the baby have grunting? \\ 
74 & During the illness that led to death did the baby have spasms or convulsions? \\ 
75 & During the illness that led to death, did the baby have fever? \\ 
76 & At what age did the fever start? \\ 
77 & How many days did the fever last? \\ 
78 & During the illness that led to death, did the baby become cold to touch? \\ 
79 & At what age did the baby start feeling cold to touch? \\
80 & How many days did the baby feel cold to touch? \\ 
\hline
\end{tabular}
\end{table}

\begin{table}[h]
\centering
\caption{List of predictors: 81-97}
\begin{tabular}{cl}
\hline
\multicolumn{1}{c}{No.}		& \multicolumn{1}{l}{Predictors}	 \\
\hline 
81 & During the illness that led to death, did the baby become lethargic after \\ & a period of normal activity? \\ 
82 & During the illness that led to death, did the baby become unresponsive or \\ & unconscious? \\ 
83 & During the illness that led to death, did the baby have a bulging fontanelle? \\ 
84 & During the illness that led to death, did the baby have pus drainage from \\ & the umbilical cord stump? \\ 
85 & During the illness that led to death, did the baby have redness of the umbilical \\ & cord stump? \\ 
86 & Did the redness of the umbilical cord stump extend onto the abdominal skin? \\ 
87 & During the illness that led to death, did the baby have skin bumps containing \\ & pus or a single large area with pus? \\
88 & During the illness that led to death, did the baby have ulcer(s) (pits)? \\
89 & During the illness that led to death, did the baby have an area(s) of skin with \\ & redness and swelling? \\
90 & During the illness that led to death, did he/she have areas of the skin that \\ & turned black? \\
91 & During the illness that led to death, did the baby bleed from anywhere? \\ 
92 & During the illness that led to death, did he/she have more frequent loose or \\ & liquid stools than usual? \\ 
93 & How many stools did the baby have on the day that diarrhea/loose liquid \\ & stools were most frequent? \\ 
94 & During the illness that led to death, did he/she vomit everything? \\ 
95 & During the illness that led to death, did he/she have yellow skin? \\ 
96 & During the illness that led to death, did the baby have yellow eyes? \\ 
97 & Did the infant appear to be healthy and then just die suddenly? \\ 
\hline
\end{tabular}
\end{table}

\clearpage


\begin{figure}[htbp]
\centering
\includegraphics[scale=0.85]{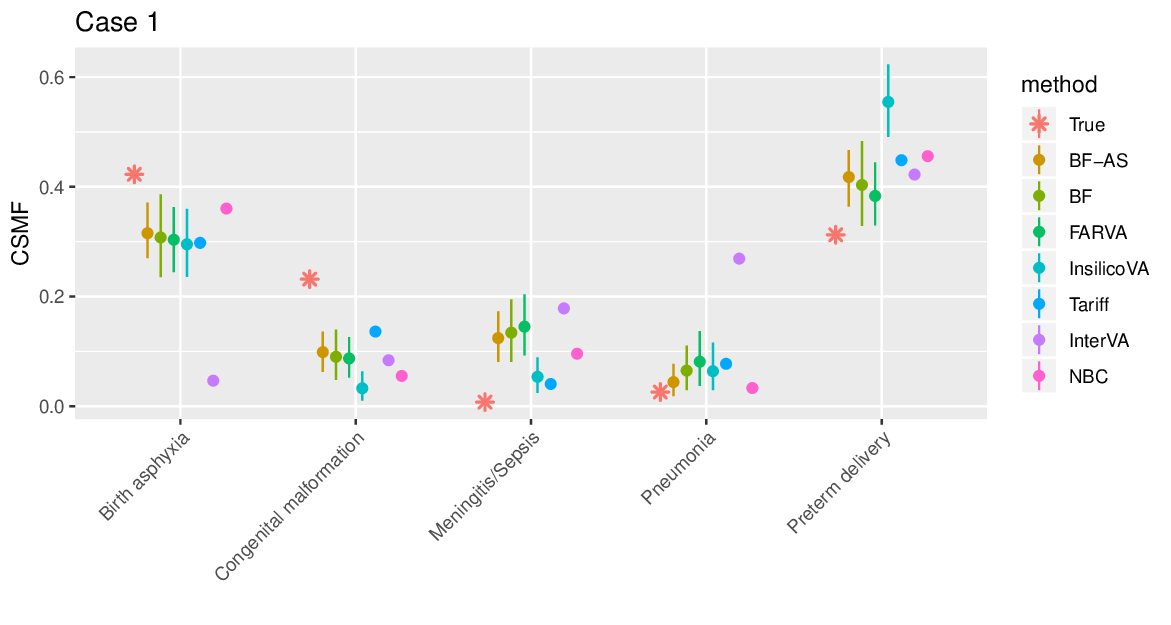}
\includegraphics[scale=0.85]{CSMF-case2-0729.eps}
\vspace{-5mm}
\caption{Estimation result of CSMF in Case 1 (above) and Case 2 (below). Red asterisk shows true values, and circle and interval correspond to mean and 95\% interval of each statistical method. BF-AS, BF, FARVA, NBC mean Bayesian factor model with age- and sex-dependence (proposed method), Bayesian factor model, Bayesian hierarchical factor regression model and Naive Bayes Classifier. Tariff, InterVA and NBC produce only the point estimates.}
\end{figure}

\begin{figure}[htbp]
\centering
\includegraphics[scale=0.85]{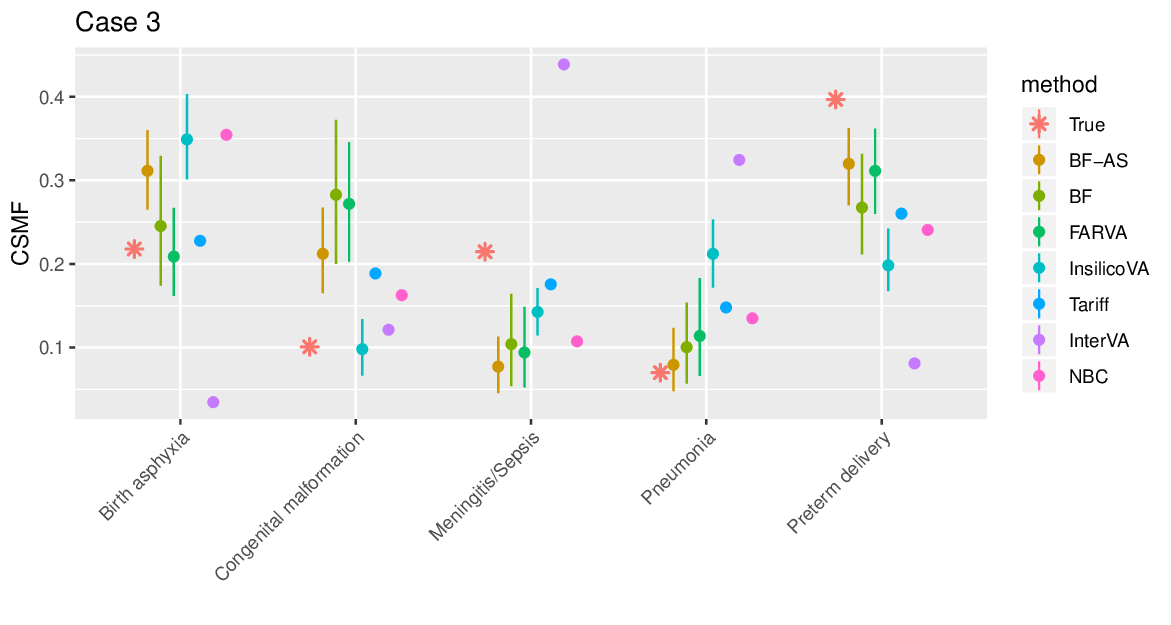}
\includegraphics[scale=0.85]{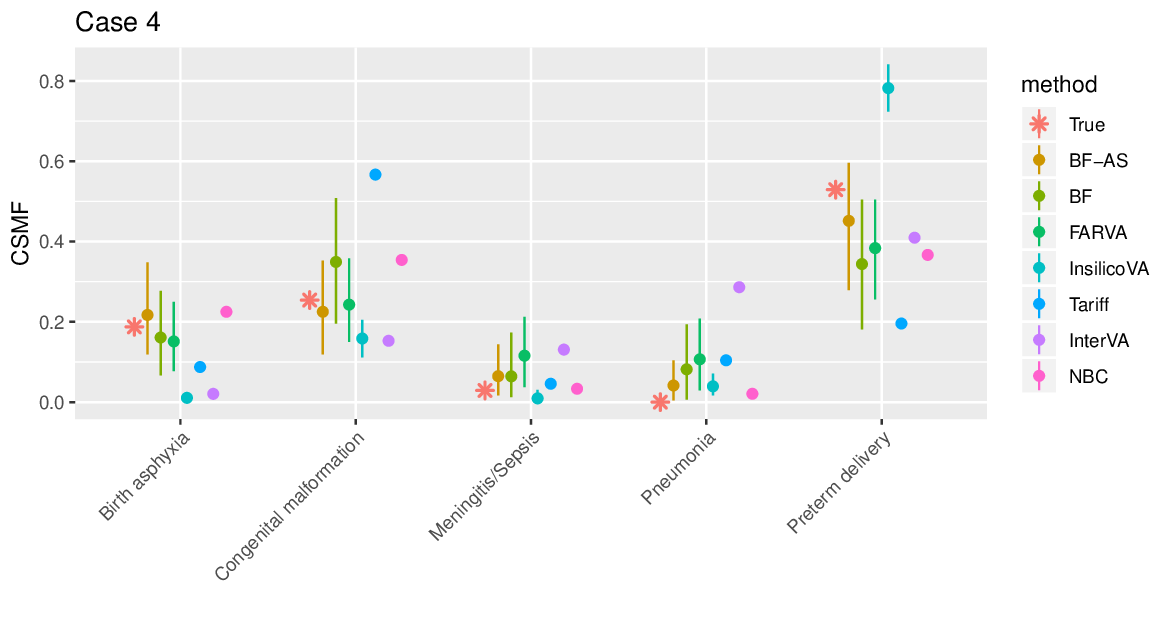}
\vspace{-5mm}
\caption{Estimation result of CSMF in Case 3 (above) and Case 4 (below). Red asterisk shows true values, and circle and interval correspond to mean and 95\% interval of each statistical method. BF-AS, BF, FARVA, NBC mean Bayesian factor model with age- and sex-dependence (proposed method), Bayesian factor model, Bayesian hierarchical factor regression model and Naive Bayes Classifier. Tariff, InterVA and NBC produce only the point estimates.}
\end{figure}

\begin{figure}[htbp]
\centering
\includegraphics[scale=0.85]{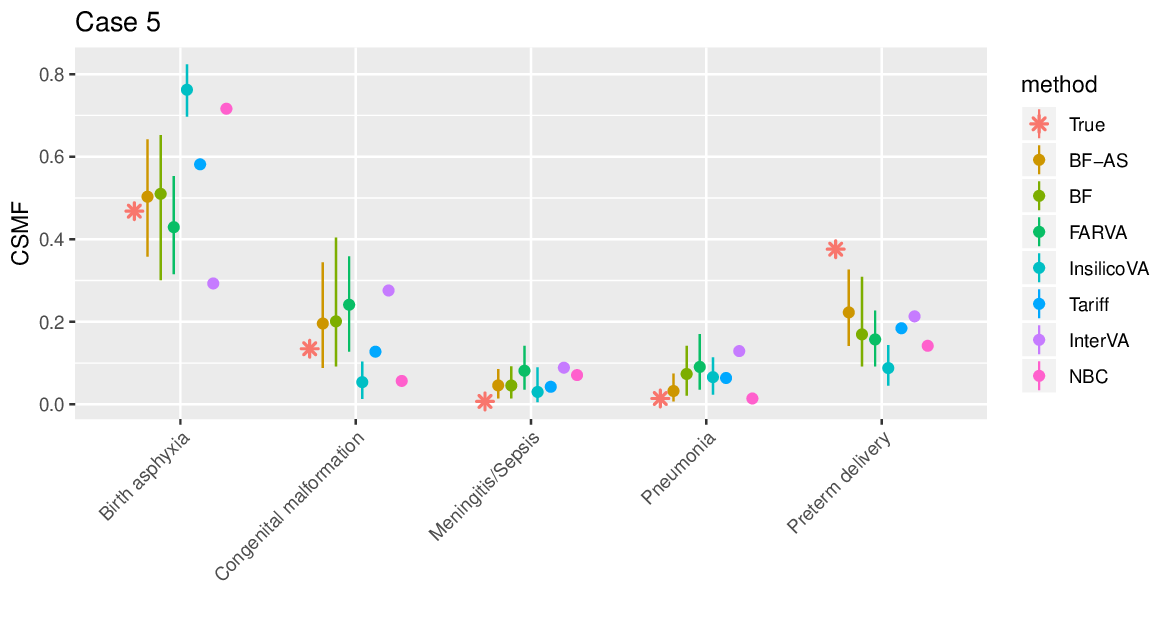}
\includegraphics[scale=0.85]{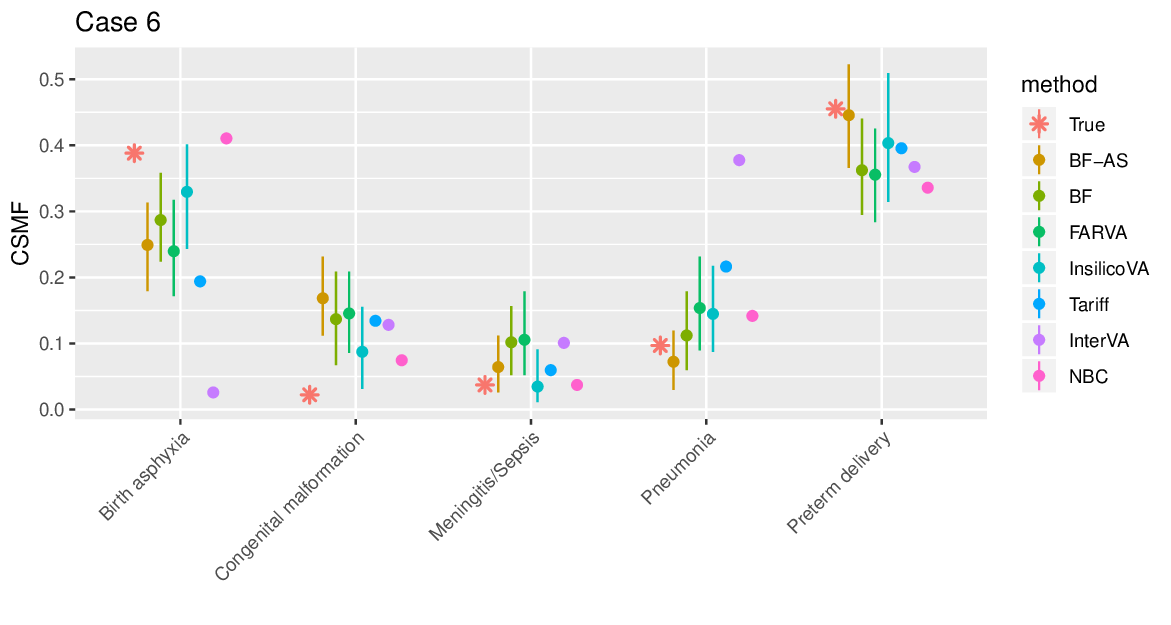}
\vspace{-5mm}
\caption{Estimation result of CSMF in Case 5 (above) and Case 6 (below). Red asterisk shows true values, and circle and interval correspond to mean and 95\% interval of each statistical method. BF-AS, BF, FARVA, NBC mean Bayesian factor model with age- and sex-dependence (proposed method), Bayesian factor model, Bayesian hierarchical factor regression model and Naive Bayes Classifier. Tariff, InterVA and NBC produce only the point estimates.}
\label{fig:case2}
\end{figure}

\clearpage


\begin{figure}[htbp]
\centering
\includegraphics[scale=0.73]{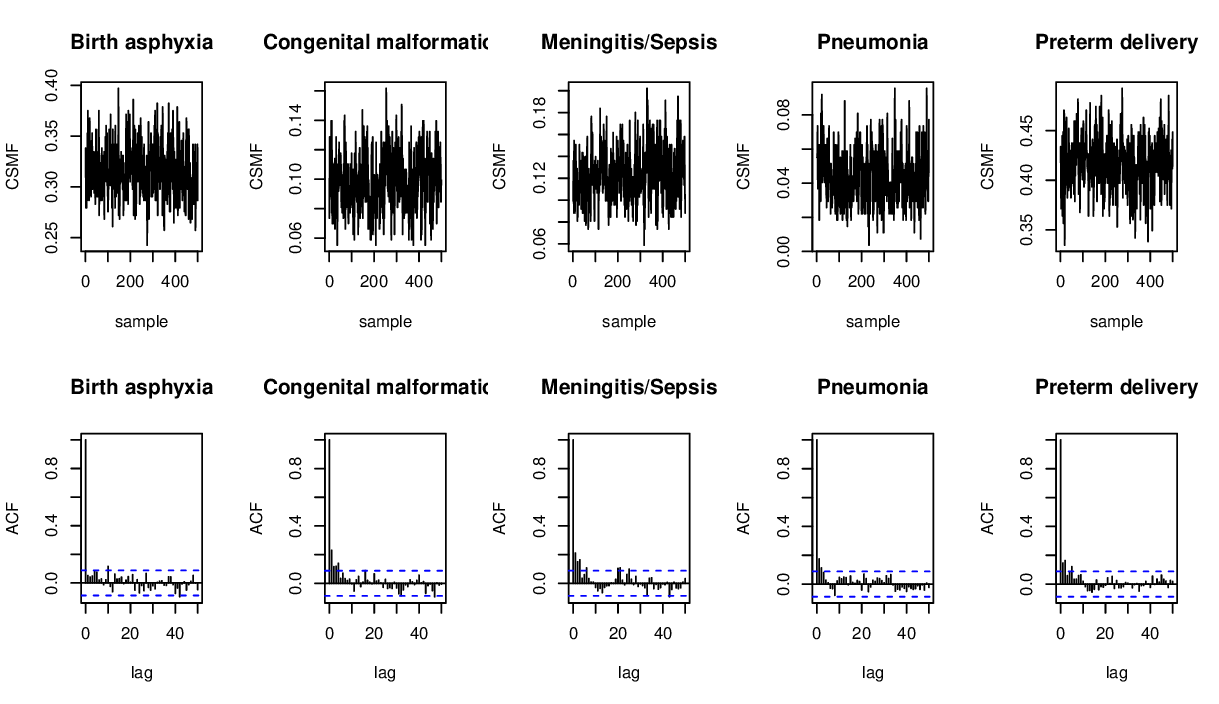}
\includegraphics[scale=0.73]{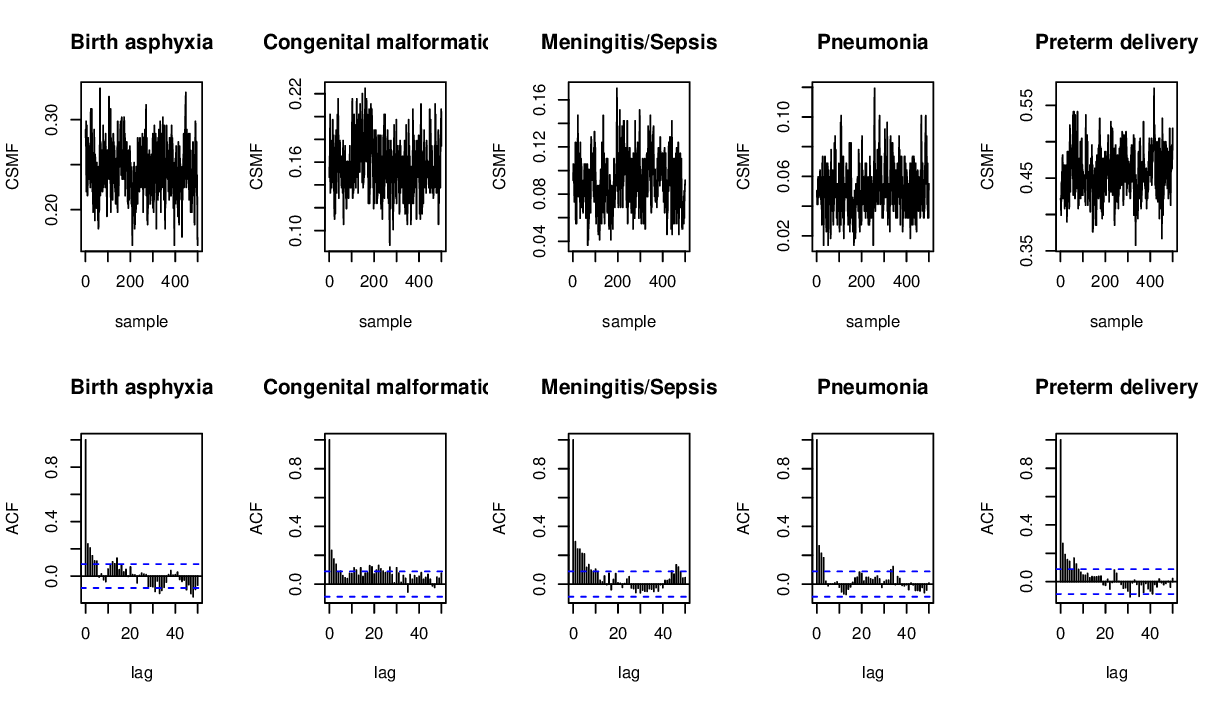}
\vspace{-5mm}
\caption{Sample paths and autocorrelations of the MCMC sample of the CSMF in the target site for the proposed model in case 1 (upper half) and case 2 (lower half).}
\end{figure}

\begin{figure}[htbp]
\centering
\includegraphics[scale=0.73]{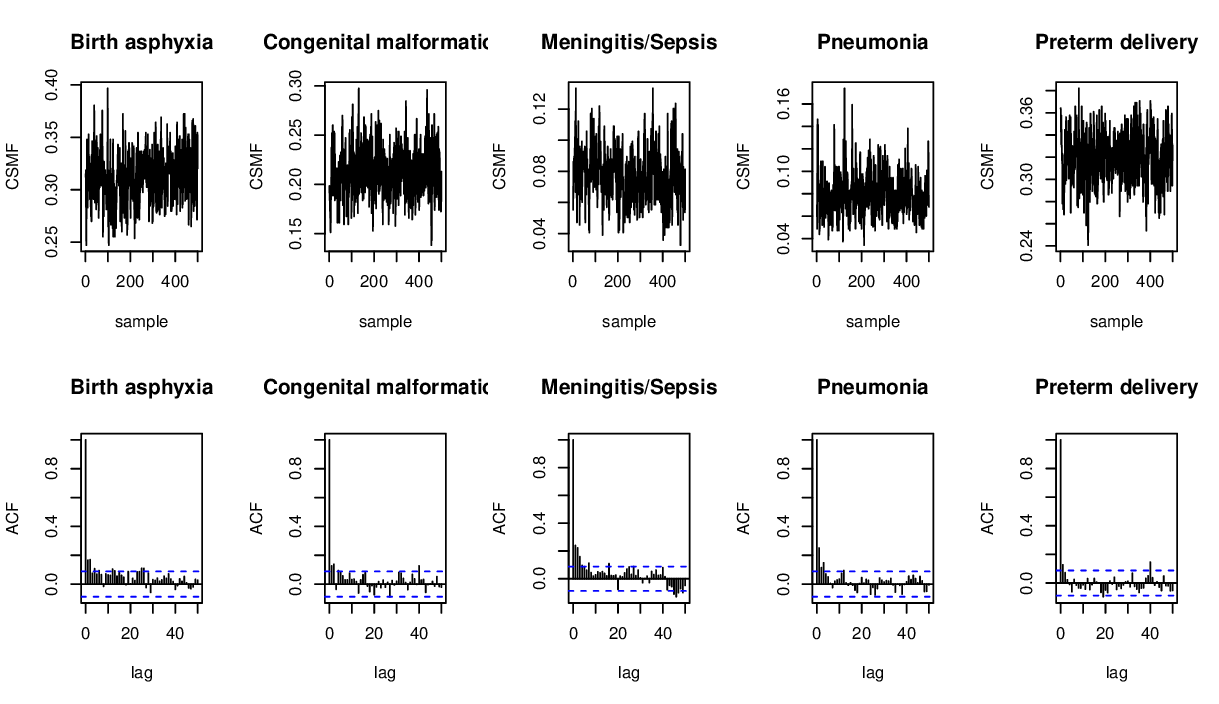}
\includegraphics[scale=0.73]{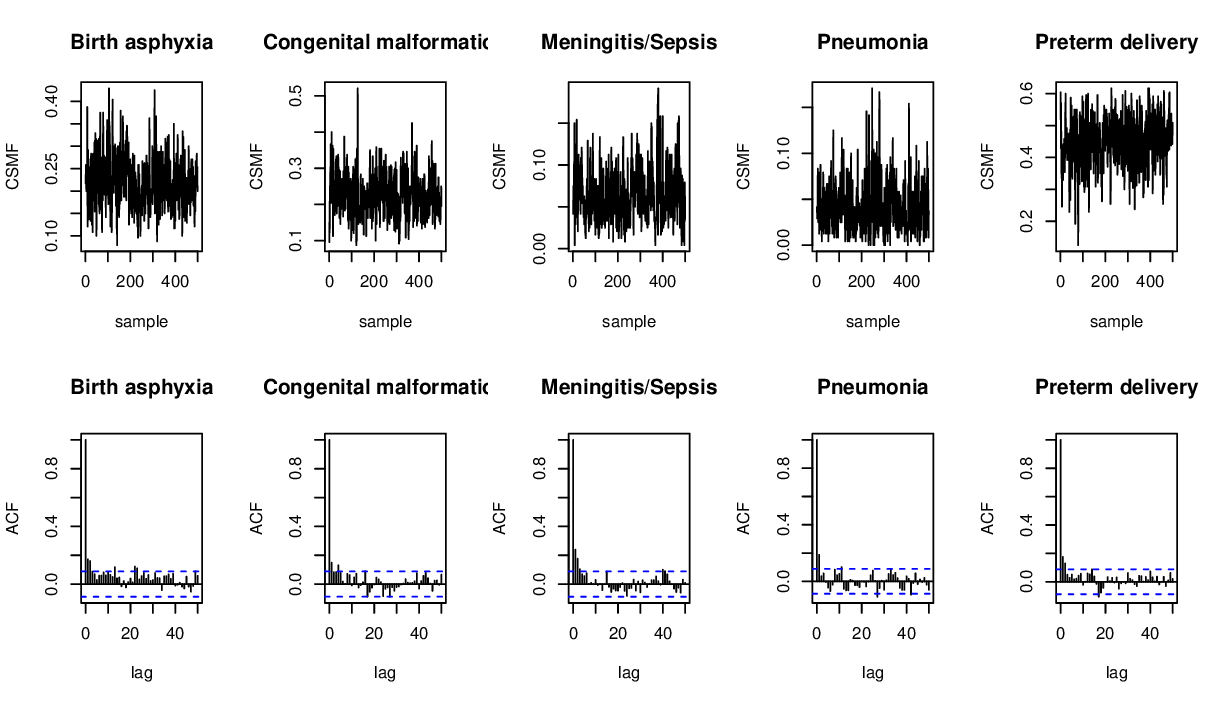}
\vspace{-5mm}
\caption{Sample paths and autocorrelations of the MCMC sample of the CSMF in the target site for the proposed model in case 3 (upper half) and case 4 (lower half).}
\end{figure}

\begin{figure}[htbp]
\centering
\includegraphics[scale=0.73]{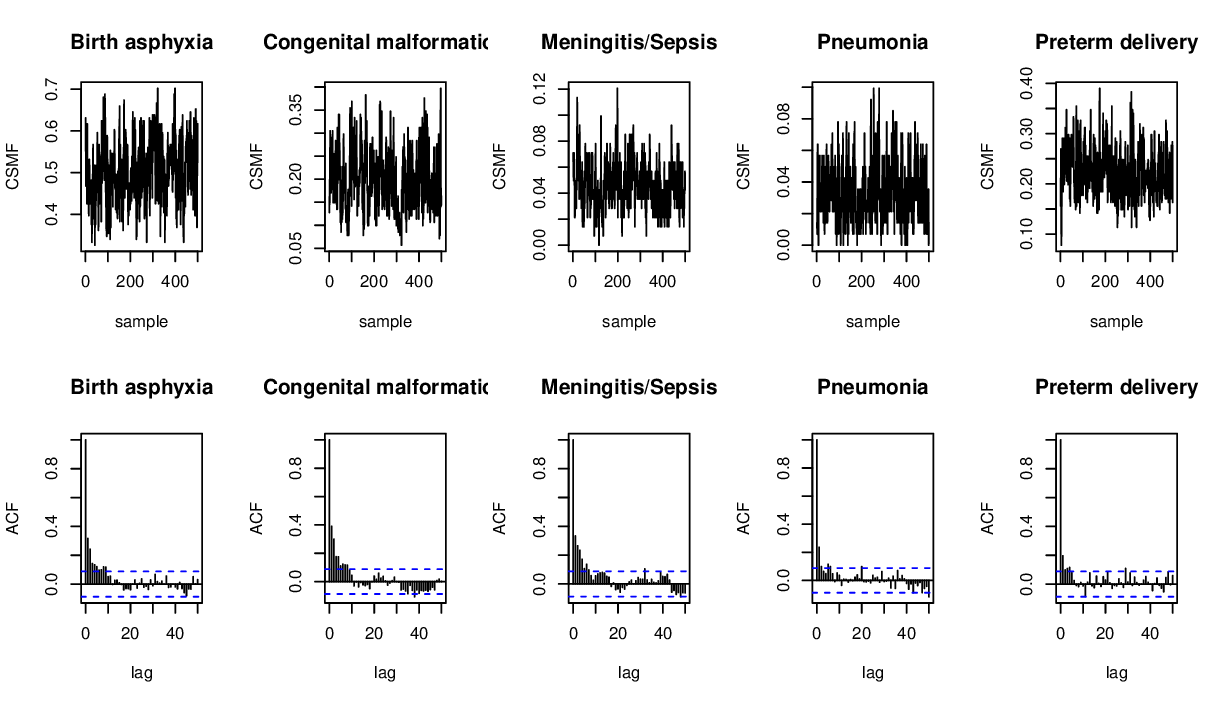}
\includegraphics[scale=0.73]{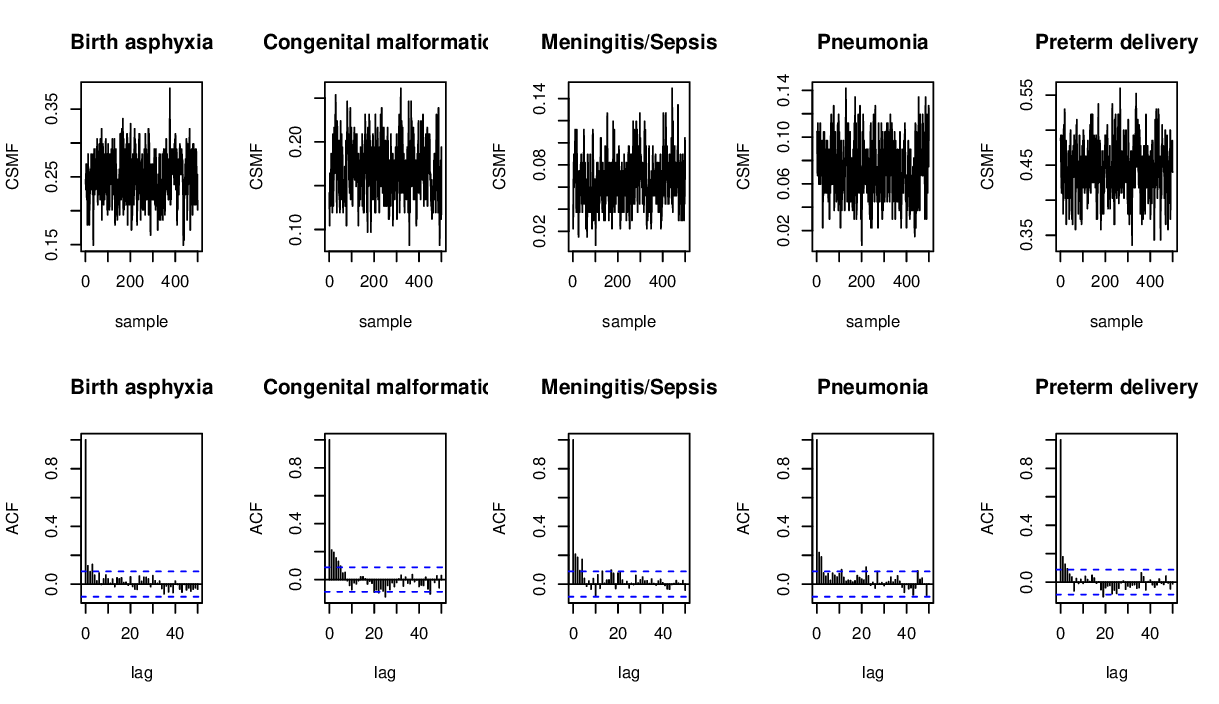}
\vspace{-5mm}
\caption{Sample paths and autocorrelations of the MCMC sample of the CSMF in the target site for the proposed model in case 5 (upper half) and case 6 (lower half).}
\label{fig:case2}
\end{figure}

\bibliographystyle{chicago}
\bibliography{VA}